\documentclass[twocolumn,prb,showpacs,floatfix,preprintnumbers]{revtex4}

\usepackage{graphicx}
\usepackage{amsfonts}
\usepackage{amsmath}
\usepackage{amssymb}

\newcommand{\ean}{\nonumber\\}
\newcommand{\ac}{\textrm{AC}}

\newcommand{\omegaL}{\omega_{\rm L}}

\newcommand{\re}{{\rm Re}}
\newcommand{\im}{{\rm Im}}

\begin{document}

\title{Comparison of classical and second quantized description of the
dynamic Stark shift}

\author{M.\ Haas}
\author{U.\ D.\ Jentschura}
\author{C.\ H.\ Keitel}

\affiliation{Max-Planck-Institut f\"{u}r Kernphysik,
Saupfercheckweg 1, 69117 Heidelberg, Germany}

\begin{abstract}
We compare the derivation of the dynamic Stark shift of hydrogenic
energy levels in  a classical framework with an adiabatically damped
laser-atom interaction, which is equivalent to the Gell-Mann-Low-Sucher
formula, and a treatment based on time-independent perturbation theory,
with a second-quantized laser-atom dipole interaction Hamiltonian. Our
analysis applies to a laser that excites a two-photon  transition in
atomic hydrogen or in a hydrogenlike ion with low nuclear charge number.
Our comparisons serve to demonstrate why  the dynamic Stark shift may be
interpreted as a stimulated radiative correction and  illustrates
connections between the two derivations. The  simplest of the
derivations is the fully quantized approach. The classical and the
second-quantized treatment are shown to be equivalent in the limit of
large photon numbers.
\end{abstract}


\maketitle

\section{Introduction}

The dynamic (AC) Stark shift is a perturbative effect that shifts atomic
energy levels in a laser field. It is an essential topic in precision
spectroscopy experiments, which have reached unprecedented
accuracy\cite{UdHoHa2002} and general
interest.\cite{SaEtAl2003,NiEtAl2000} The dynamic Stark effect is well
within the reach of students who have studied the quantum mechanics of
the hydrogen atom and can be understood at different levels in 
theoretical physics courses.

The two approaches we will present, the classical field and fully
quantized field description, coincide in the classical limit of high
photon density, as is to be expected. So the AC Stark effect also serves
as an example of how to connect and contrast classical and quantum
notions of physical phenomena.

If an atom is exposed to external electromagnetic fields, its energy
levels are shifted due to the interaction of the electrons with the
field. This shift of energy levels can be observed in spectroscopic
experiments, where for example, absorption or fluorescence spectra are
measured. The Zeeman effect describes this energy shift for static
magnetic fields, and the DC Stark effect is responsible for the level
shift in static electric fields. Both these effects can be avoided in
principle by a proper shielding of the atom. However, the probing laser
light with which atoms are irradiated in order to obtain a spectrum also
constitutes a time-dependent electromagnetic field and is necessarily
always present in laser spectroscopy. Its impact on  atomic energy
levels is called the AC Stark effect and for nondegenerate states it can
be understood as a time averaged DC Stark shift, as explained in the
Appendix. This  statement holds only for off-resonant driving of the
atom, where the AC Stark shift can be considered as a perturbation.

In this article off-resonant driving is to be understood with regard to
any electric-dipole allowed, one-photon transition. Even in the
nonresonant case two-photon transitions can be driven effectively when
the frequency of the incident radiation is close to half the atomic
transition frequency. 

Before we discuss  the off-resonant excitation of an atom by laser
radiation, we briefly mention the main differences with the resonant
case.\cite{brief}  Most importantly, if the frequency of the incident
radiation is close to a one-photon resonance, then the influence of the
laser field on the atomic levels lies outside the regime of perturbation
theory and must be included nonperturbatively in the dressed-state
picture. The reason is that the dipole matrix element between the states
involved is nonzero for one-photon transitions. Consequently,  the level
shift is linear in the electric field amplitude of the laser, in
contrast to the quadratic dependence that we will obtain for
off-resonant excitation. For further information on dressed states we
refer the reader to 
Refs.~\onlinecite{Mo1969,JeEvHaKe2003,JeEvKe2005,EvJeKe2004}.

In a classical framework the dynamic Stark shift can be described by
time-dependent perturbation theory. We will demonstrate that the dynamic
Stark shift can  be used to illustrate some basic aspects of quantum
electrodynamics (QED). The dominant shift of the energy levels in this
case can be attributed to a second-order perturbation  in which a
laser-photon is created or annihilated in a virtual intermediate state. 

This article is also devoted to showing that the AC Stark shift can be
identified as a stimulated radiative correction.\cite{CTDRGr1992}
Indeed, the AC Stark shift is approximately equivalent to a spectral
component of the electron self-energy (Lamb shift) that results when we
restrict the discussion to a single mode of the electromagnetic field,
but with an important difference: for the Lamb shift the photon modes
are all unoccupied in the unperturbed state in contrast to the AC Stark
shift for which there is  one highly occupied mode of the
electromagnetic field, the laser mode. 

This paper is organized as follows: In Sec.~\ref{classical} the dynamic
Stark shift is described using a classical laser field, which
necessitates the use of time-dependent perturbation theory and  an
adiabatically damped interaction. In contrast in Sec.~\ref{quantized} we
derive the dynamic Stark shift using a quantized-field approach and
time-independent perturbation theory. The classical and second-quantized
results are shown to agree in the classical limit, that is, for a
macroscopically populated laser field mode. 

\section{Classical Field Approach}
\label{classical}

In this section we rederive the classical expressions for the dynamic
(AC) Stark shift using a classical description of the laser field. Our
approach  is the usual one employed in the literature, and our treatment
and our notation is inspired by Chap.~5 of Ref.~\onlinecite{Sa1994Mod}.
Let us consider the Hamiltonian
\begin{equation}
H=H_0+V(\epsilon,t), \label{eq:H}
\end{equation} 
where
\begin{subequations}
\begin{align}
\label{defH0}
H_0&=\frac{{\bf p}^2}{2m_{\rm e}} - 
\frac{Z e^2}{4\pi\epsilon_0 r}\,,\\ 
\label{defVeps}
V(\epsilon, t)&=V\exp(- \epsilon |t|)
\cos (\omega_{\rm L} t),\\
\label{defV}
V&=-e z {\cal E}_{\rm L}.
\end{align}
\end{subequations}
The Hamiltonian in Eq.~\eqref{eq:H} describes a hydrogen atom ($Z=1$) or
a hydrogen-like ion of nuclear charge number $Z>1$ in a plane-wave
monochromatic laser field, polarized along the $z$-direction,
adiabatically damped in the distant past ($t \to -\infty$) and the
distant future ($t \to \infty$). $H_0$ describes the nonrelativistic
unperturbed hydrogen Hamiltonian, and $V(\epsilon, t)$ is the
time-dependent, adiabatically damped, harmonic perturbation with
magnitude $V$; $\epsilon$ is the infinitesimal damping parameter (see,
for example, p.~342 of Ref.~\onlinecite{Sa1994Mod}). We have assumed
that the wavelength of the driving light of angular frequency
$\omega_{\rm L}$ is large compared to the spatial extent of the atomic
wave functions (the dipole approximation). The laser-atom interaction
$V(\epsilon, t)$ is treated in the length gauge as in
Ref.~\onlinecite{Ko1983} with electric field amplitude ${\cal E}_{\rm
L}$. The electric field strength involved in $V(\epsilon, t)$ is a
gauge-invariant quantity.\cite{Ko1983}

The  parameter $\epsilon>0$ is introduced to avoid a sudden turn-on of
the perturbation. In the limit  $\epsilon \to 0$ we will obtain the
constant intensity result after carrying out the relevant time
integrations of the first few terms in the Dyson series. The
introduction of an adiabatic damping parameter is also a key element of
time-dependent perturbation theory in QED.\cite{MoPlSo1998} In QED the
interaction Hamiltonian  is usually expressed in the interaction picture
and a time-dependence is incurred for the field operators (see
Appendix~A of Ref.~\onlinecite{JeKe2004aop}). 

Energy shifts in QED are usually formulated using the
Gell-Mann-Low-Sucher theorem.\cite{GMLo1951,Su1957} The applicability of
this theorem is not restricted to the case of perturbations in a
second-quantized approach, but can be applied equally well to a
time-dependent, classical perturbation.

We now consider the effect of the off-resonant perturbation by an
time-dependent electric field on a reference state $|\phi\rangle$ of the
unperturbed atom. In the interaction picture (denoted by the subscript
$I$), $V(\epsilon,t)$ is represented by 
\begin{equation}
V_{\rm I}(\epsilon, t)=
\exp\Big(\frac{i}{\hbar}H_0 t\Big) V(\epsilon,t)
\exp\Big(-\frac{i}{\hbar} H_0 t\Big).
\end{equation}
From the Dyson series we can calculate the time evolution operator
$U_{\rm I}(\epsilon,t)$ up to second order in $V_{\rm I}$:
\begin{eqnarray}
\label{eqn:Dyson}	 
U_{\rm I}(\epsilon, t)&=&
1-\frac{i}{\hbar} 
\!\int\limits_{-\infty}^t dt' V_{\rm I}(\epsilon, t')
\ean
&&+ \Big(-\frac{i}{\hbar}\Big)^2
\!\int\limits_{-\infty}^t dt'
\!\int\limits_{-\infty}^{t'} dt'' 
V_{\rm I}(\epsilon, t') 
V_{\rm I}(\epsilon, t'').
\end{eqnarray}
Now consider the time-dependent atomic state $| \psi_{\rm I}(t) \rangle$
in the interaction picture subject to the initial condition  $|\psi_{\rm
I}(t\!=\!-\infty)\rangle=|\phi\rangle$,  where the reference state
$|\phi\rangle$ is an eigenstate of the unperturbed Hamiltonian $H_0$. We
expand $| \psi_{\rm I}(t) \rangle$ in a complete set $\{|m\rangle\}$ of
eigenstates of $H_0$ as
\begin{equation}
\label{psiI}
| \psi_{\rm I}(t) \rangle = 
U_{\rm I}(\epsilon, t) | \psi_{\rm I}(-\infty)\rangle = 
\sum_m c_m(t) | m \rangle,
\end{equation}
where $c_m(t)=\langle m|\psi_I(t)\rangle$. The initial condition is thus
$c_\phi(-\infty) = 1$ for the reference state $|\phi\rangle$ with all
other $c_m(-\infty)$ equal to zero. We are interested in the projection 
\begin{equation}
\label{cphi}
c_\phi(t)= \langle \phi | \psi_{\rm I}(t) \rangle = 
\langle \phi |U_{\rm I}(\epsilon, t)|\phi \rangle.
\end{equation}
We substitute $U_{\rm I}(\epsilon, t)$ from Eq.~(\ref{eqn:Dyson}) and
because $\langle \phi|z|\phi\rangle$ vanishes for parity eigenstates
$|\phi\rangle$, the leading order is $V^2$ and the problem reduces to
calculating the matrix element
\begin{subequations}
\begin{align}
\label{resM}
M & = \!\int\limits_{-\infty}^t dt'
\!\int\limits_{-\infty}^{t'} dt''
\langle \phi|V_{\rm I}(\epsilon, t') 
V_{\rm I}(\epsilon, t'')|\phi\rangle \\
& = \sum\limits_m \!\int\limits_{-\infty}^t dt'
\!\int\limits_{-\infty}^{t'} dt''
\langle \phi|V_{\rm I}(\epsilon, t')|m\rangle
\langle m| V_{\rm I}(\epsilon, t'')|\phi\rangle,
\end{align}
\end{subequations}
where the multi-index $m$ counts all bound and continuum states of the
unperturbed hydrogen atom. Because the perturbation is harmonic, the
time integrals can be done  without  difficulty, convergence being
ensured by the adiabatic damping. We obtain
\begin{equation}
\label{Mintegrated}
M=-\frac{\hbar}{i}\frac{1}{4}\sum\limits_{m,\pm}
\frac{\langle\phi|V|m\rangle\langle m|V|\phi\rangle
\exp(2\epsilon t)}
{2\epsilon (E_\phi-E_m \pm\hbar \omega_{\rm L} + 
i \hbar \epsilon)},
\end{equation}
with $V$ as defined in Eq.~(\ref{defV}); $E_\phi$ represents the energy
of the unperturbed atomic state $|\phi\rangle$. The $\pm$ index denotes
the summation of the two terms differing only in the sign of
$\hbar\omega_{\rm L}$ in the denominator. This sum and the factor of
$1/4$ originate from the definition of the cosine in terms of
exponential functions. In view of Eqs.~(\ref{eqn:Dyson}),~(\ref{cphi}),
and (\ref{Mintegrated}), we have in second-order time-dependent
perturbation theory
\begin{equation}
\label{cphires}
c_\phi(t)=1-\frac{i}{4 \hbar}\sum\limits_{m,\pm}
\frac{\langle\phi|V|m\rangle\langle m|V|\phi\rangle
\exp(2\epsilon t)}
{2\epsilon (E_\phi-E_m\pm\hbar\omega_{\rm L}+i\hbar\epsilon)} 
+ \dots,
\end{equation}
where higher-order terms have been  neglected. Now consider
\begin{equation}
\label{diffeq}
\frac{\partial}{\partial t}\ln(c_\phi(t))=
-\frac{i}{4 \hbar}\sum_{m,\pm}
\frac{\langle \phi|V|m\rangle\langle m|V|\phi\rangle}
{E_\phi-E_m\pm\hbar\omega_{\rm L}+i \hbar\epsilon}.
\end{equation}
Here the logarithm has been expanded up to second order in $V$ and
$\exp(2 \epsilon t)$ has been replaced by unity. The solution of
Eq.~(\ref{diffeq}) implies that 
\begin{equation}
c_\phi(t) =
\exp\big(-\frac{i}{\hbar} \Delta E_{\ac}(\phi) t\big),
\end{equation}
where we have defined the dynamic Stark shift $\Delta E_{\ac}(\phi)$ of
the reference state $|\phi\rangle$
\begin{equation}
\label{defAC}
\Delta E_{\ac}(\phi)=\frac{1}{4}\sum_{m,\pm}
\frac{\langle \phi|V|m\rangle\langle m|V|\phi\rangle}
{E_\phi-E_m\pm\hbar\omega_{\rm L}+ i\hbar\epsilon}.
\end{equation}
In view of Eq.~(\ref{psiI}), we have
\begin{equation}
\label{PsiIResult}
| \psi_{\rm I}(t) \rangle = 
U_{\rm I}(\epsilon, t) | \psi_{\rm I}(-\infty) \rangle = 
c_\phi(t) | \phi \rangle + \dots,
\end{equation}
where the ellipsis denotes the projections onto the nonreference atomic
states. Because the Schr\"odinger picture wave function is related to
its interaction-picture counterpart via $| \psi(t) \rangle = \exp(-i H_0
t) | \psi_{\rm I}(t) \rangle$, we  have
\begin{equation}
\label{finalproject}
\langle \phi | \psi(t) \rangle = 
\exp\big(-\frac{i}{\hbar}
(E_\phi+\Delta E_{\ac}(\phi))t\big).
\end{equation}
The nonreference states from Eq.~(\ref{PsiIResult}) give no contribution
because they are orthogonal to $|\phi\rangle$.  The projection
(\ref{finalproject}) yields the influence of the perturbation on the
reference state $|\phi\rangle$ by projecting the time-evolved perturbed
state onto the reference state such that the perturbation to the
eigenenergy $E_\phi$ can be directly seen.

Note that $\Delta E_{\ac}(\phi)$ can in general be complex, rather than
real. We define
\begin{align}
\gamma_\phi &= - \frac{2}{\hbar}
\im(\Delta E_{\ac}(\phi)),
\label{eqn:amplexplicit}\\
\Delta E_\phi &= \re(\Delta E_{\ac}(\phi)).
\end{align}
The real part of the AC Stark effect describes the energy shift of the
unperturbed energy $E_\phi$, and the imaginary part, if present, can be
interpreted as the ionization rate $\gamma_\phi$. We can now express the
dynamic Stark shift of $|\phi\rangle$ as 
\begin{equation}
\label{greenAC}
\Delta E_{\ac}(\phi) =
\frac{1}{4} \sum_{\pm}
\Big< \phi \Big| 
V\frac{1}{E_\phi-H_0\pm
\hbar\omega_{\rm L}+i\hbar\epsilon}V
\Big| \phi \Big> ,
\end{equation}
where the closure relation for the spectrum is employed. In the Appendix
the zero-frequency limit of Eq.~\eqref{greenAC} is related to the static
Stark effect. Equation~(\ref{greenAC}) can be written conveniently as a
product of a prefactor and a sum of two matrix elements, where $E$ is
the energy of the respective intermediate state \mbox{$E = E_\phi\mp
\hbar\omega_{\rm L}$}:
\begin{subequations}
\label{defEACclas}
\begin{align}
P_{\omega_{\rm L}}(\phi)&=
\sum\limits_\pm
\Big< \phi \Big|z\frac{1}{H_0-E_\phi\pm\hbar\omega_{\rm L}}z
\Big|\phi\Big>,
\label{eqn:pomega}\\
\Delta E_{\ac}(\phi)&=
-\frac{e^2 {\cal E}_{\rm L}^2}{4} P_{\omega_{\rm L}}(\phi) =
-\frac{e^2}{2 c \epsilon_0} I P_{\omega_{\rm L}}(\phi),
\label{eqn:EACclass}
\end{align}
\end{subequations}
where $P_{\omega_{\rm L}}(\phi)$ is the dynamic polarizability of the
atom in the reference state for angular frequency $\omega_{\rm L}$ of
the driving laser field. The intensity $I$ of a plane electromagnetic
wave is
\begin{equation}
I=\frac{1}{2}\epsilon_0 c {\cal E}_{\rm L}^2.
\end{equation}
This derivation completes our analysis of the AC Stark shift using an
adiabatically damped,\cite{Sa1994Mod} classical-field\cite{Ko1983}
approach.

To illustrate the connection to the Gell-Mann-Low-Sucher theorem, we
observe that $c_\phi(0)$ in Eqs.~(\ref{cphi}) and (\ref{cphires}) can be
identified with the quantity $\langle \alpha | U(0, -\infty;\epsilon) |
\alpha \rangle$ in the notation of Ref.~\onlinecite{Su1957}, and the
perturbation $V(\epsilon, t)$ as defined in Eq.~(\ref{defVeps}) has to
be supplemented by an auxiliary scaling variable $g$,
\begin{equation}
V(\epsilon, t) \to g V(\epsilon, t).
\end{equation}
The parameter $g$ is later set equal to unity. We then have according to
Eq.~(2$\pm$) of Ref.~\onlinecite{Su1957},
\begin{subequations}
\begin{align}
\label{trafo}
\Delta E &= \lim_{g \to 1}
\Big( i \hbar\epsilon\,g
\frac{\partial}{\partial g}
\ln \big< \phi | U_{\rm I}(\epsilon, 0)| 
\phi \big> \Big) \\
&= \lim_{g \to 1}
\Big( i \hbar\epsilon\,g 
\frac{\partial}{\partial g}
\ln \Big( 1 - g^2\frac{M}{\hbar^2} \Big) \Big) \\
&\approx \lim_{g \to 1} 
\frac{g}{8} \frac{\partial}{\partial g}
\Big(\sum_{m,\pm}
\frac{\langle \phi|g V|m\rangle\langle m|g V|\phi\rangle}
{E_\phi-E_m\pm\hbar\omega_{\rm L}+i \hbar\epsilon}\Big) \label{trafo.c} \\
&= \frac{1}{4} 
\sum_{m,\pm}
\frac{\langle \phi|V|m\rangle\langle m|V|\phi\rangle}
{E_\phi-E_m\pm\hbar\omega_{\rm L}+i \hbar\epsilon}.
\end{align}
\end{subequations}
The latter result $\Delta E = \Delta E_{\rm AC}(\phi)$ agrees with
Eq.~(\ref{defAC}). In the step leading to Eq.~(\ref{trafo.c}), an
expansion of the logarithm in powers of $g$ is implied, which is
equivalent to a second-order expansion in the time-dependent
perturbation $V$.

\section{Fully Quantized Approach}
\label{quantized}

In the classical picture we set the field amplitude to a constant value
${\cal E}_{\rm L}$ and used time-dependent perturbation theory with an
adiabatic damping parameter. When treating the light as a photon field,
the classical picture can be interpreted as the limit of the fully
quantized treatment in the limit of a large photon number.

In second quantization the Hamiltonian for the coupled system, atom $+$
radiation field, reads
\begin{equation}
\label{H0AF}
H=\int\hspace{-0.5cm}\sum\limits_{n} E_n |n\rangle\langle n| + 
\hbar \omega_{\rm L} a_{\rm L}^\dag a_{\rm L} 
+ H_{\rm L}.
\end{equation}
The first term contains a sum over the discrete spectrum and an integral
over the continuous spectrum of the Schr\"{o}dinger equation. We do not
consider electron-positron pair creation, and therefore we do not
quantize the fermion field. The laser field is described as a quantized
photon field with creation and annihilation operators $a_{\rm L}^\dag$
and $a_{\rm L}$, respectively. $H_{\rm L}$ reads (in the length gauge)
\begin{equation}
H_{\rm L} = -e z \hat{E}_{\rm L}
= -e z \sqrt{\frac{\hbar \omega_{\rm L}}
{2 \epsilon_0 {\cal V}}}
(a_{\rm L} + a_{\rm L}^\dag).
\end{equation}
(See also Eqs.~(4.7) and~(4.8) of Ref.~\onlinecite{JeKe2004aop}.) The
symbol ${\cal V}$ denotes the normalization volume and is chosen so that
the energy density of a one-photon Fock state when integrated over
${\cal V}$ yields   $\hbar \omega_{\rm L}$.  It might be argued that a
coherent state  of the photon field is a much better description than a
Fock state with $n_{\rm L}$ photons in the laser mode, which we have
assumed here. However, in the limit of large photon number, the relative
fluctuation of the photon number $\delta n_{\rm L}/n_{\rm L}$ goes to
zero for a coherent state, and we may therefore resort to the Fock-state
approximation.\cite{see}

We work in the Schr\"odinger picture where the field operators carry no
time dependence. It is not so widely known that it is possible to
formulate time-independent operators for the quantized radiation field,
let alone to do meaningful calculations with these operators. However,
this formulation is introduced in a few textbooks such as 
Ref.~\onlinecite{KuSt1995}.

The concept of time-independent field operators has also been used for
quantum electrodynamic calculations (see for example, Eq.~(5) of
Ref.~\onlinecite{Pa1998}). Following this approach, we are now in a
position to apply time-independent perturbation
theory.\cite{JeKe2004aop} This approach leads to the following
second-order result for the energy shift of the unperturbed eigenstate
$|\phi,n_L\rangle$, 
\begin{align}
\label{eqn:EAC1}
\Delta E_{\ac}(\phi) 
&= \nonumber \\
\sum\limits_m
\Big [&
\frac{\langle \phi,n_{\rm L}|H_{\rm L}|m,n_{\rm L}-1\rangle 
\langle m,n_{\rm L}-1|H_{\rm L}|\phi,n_{\rm L}\rangle}
{E_\phi+ n_{\rm L} \hbar \omega_{\rm L}-
(E_m+(n_{\rm L}-1) \hbar \omega_{\rm L})} \nonumber \\
&+\frac{\langle \phi,n_{\rm L}|H_{\rm L}|m,n_{\rm L}+1\rangle 
\langle m,n_{\rm L}+1|H_{\rm L}|\phi,n_{\rm L}\rangle}
{E_\phi+ n_{\rm L}\hbar\omega_{\rm L}-
(E_m+(n_{\rm L}+1)\hbar\omega_{\rm L})}
\Big] \nonumber \\
&=
\frac{e^2\hbar\omega_{\rm L}}{2\epsilon_0{\cal V}}
\sum\limits_m
\Big[
\frac{\langle \phi|z|m\rangle \langle m|z|\phi\rangle}
{E_\phi-E_m+\hbar \omega_{\rm L}}n_{\rm L}\nonumber \\
&{}\quad +\frac{\langle \phi|z|m\rangle \langle m|z|\phi\rangle}
{E_\phi-E_m-\hbar \omega_{\rm L}}(n_{\rm L}+1)
\Big].
\end{align}
The sum over virtual intermediate states $|m\rangle$ has been split into
two parts depending on the number of photons in the field. In the
classical limit $n_{\rm L} \to \infty$, ${\cal V} \to \infty$, $n_{\rm
L}/{\cal V}={\rm constant}$, we can simplify Eq.~(\ref{eqn:EAC1}) to
read
\begin{equation}
\label{eqn:deEACquantum}
\Delta E_{\ac}(\phi)=
-\frac{e^2\hbar n_{\rm L}\omega_{\rm L}}
{2\epsilon_0{\cal V}} P_{\omega_{\rm L}}(\phi),
\end{equation}
with $P_{\omega_{\rm L}}(\phi)$ as given in Eq.~(\ref{eqn:pomega}).

The remaining issue concerns the matching of this result to the
classical result in Eq.~(\ref{defEACclas}). In the quantized formalism
the term
\begin{equation}
\label{help1}
w=\frac{n_{\rm L}\hbar\omega_{\rm L}}{{\cal V}}
\end{equation}
gives the energy density in which the atom is immersed, which is related
to the intensity via
\begin{equation} 
\label{help2}
I=wc.
\end{equation} 
We use Eqs.~(\ref{eqn:deEACquantum})--(\ref{help2}) and obtain
\begin{equation}
\Delta E_{\ac}(\phi)=
-\frac{e^2}{2\epsilon_0c}IP_{\omega_{\rm L}}(\phi),
\end{equation}
in agreement with Eq.~(\ref{defEACclas}). Thus the classical-field and
the quantized-field approach give consistent results in the classical
limit. 

\section{Conclusions}
\label{conclu}

We have contrasted two ways of deriving  analytic expressions for the
dynamic Stark shift of a hydrogenic energy level. The first, based on an
adiabatically damped length-gauge interaction (see Eq.~(\ref{defVeps})),
leads to a classical treatment where the electric laser field is simply
modeled as a periodic perturbation (see Sec.~\ref{classical}). The
second derivation, based on a quantized description of the
electromagnetic field, leads to expressions that are equivalent to the
classical expressions in the limit of a large occupation number of the
laser mode~(see Sec.~\ref{quantized}). 

The AC Stark shift has been characterized as a stimulated radiative
correction\cite{CTDRGr1992} because it results from a self-energy-like
formalism if the sum over virtual modes of the photon field is
restricted to a single mode: the laser mode. We illustrated this
statement by giving an explicit derivation in Eq.~(\ref{eqn:EAC1}). This
treatment is  based on time-independent field operators. Equation
(\ref{eqn:EAC1}) illustrates how the classical predictions should be
modified in an environment where the photon number is not large. Indeed,
the AC Stark shift receives an interpretation in this context as the
second-order perturbation incurred by the coupled system, atom + laser
field, due to virtual creation and annihilation of laser photons. When
the perturbation is evaluated using an empty Fock space as the
unperturbed state, and when a sum is formed over all possible virtual
excitations, the self-energy is obtained.\cite{JeKe2004aop,Ho2004}.

\section*{Acknowledgements}

The authors acknowledge helpful conversations with N.\ Kolachevsky, 
V.\ Yakhontov, J.\ Evers, and P.\ J.\ Mohr.  U.D.J. acknowledges support
from the Deutsche Forschungsgemeinschaft  via the Heisenberg program.

\clearpage

\appendix

\section{DC Stark shift}

For nondegenerate states, the DC Stark shift is a  second-order
perturbation  in the electric field strength and can be interpreted as
the zero-frequency limit of the AC Stark shift. In this appendix we
briefly illustrate this relation. In nonrelativistic quantum mechanics
the ground state is the only nondegenerate state. However, as a
consequence of the spin of the electron (fine structure), the spin of
the nucleus (hyperfine structure) and QED effects (Lamb shift), the
degeneracy of other states is broken, and as long as the DC Stark shift
is small compared to these energy differences, the following
considerations also hold for excited states. For larger perturbations,
the DC Stark effect is linear in the electric field. 

For a state $|\phi\rangle$ that fulfills the above conditions, consider
the limit of the dynamic Stark shift obtained in Eq.~(\ref{defEACclas})
as the angular frequency of the laser field goes to zero
\begin{align}
\label{AClimitDC}
\Delta E_{\rm AC,0}=\lim\limits_{\omegaL \to 0}
-\frac{e^2{\cal E}_{\rm L}^2}{4}
\sum\limits_\pm
\Big\langle \phi \Big|z\frac{1}{H_0-E_\phi\pm\hbar\omegaL}
z\Big|\phi\Big\rangle.
\end{align}
We use the relation
\begin{align}
&\lim_{\eta \to 0} \frac12 \Big(
\frac{1}{H_0-E_\phi+\eta}+
\frac{1}{H_0-E_\phi-\eta}\Big)
=\Big(\frac{1}{H_0-E_\phi}\Big)'\nonumber \\
&=\int\hspace{-0.55cm} \sum\limits_{m\neq\phi}
\frac{|m\rangle\langle m|}{E_m-E_\phi},
\label{redGreen}
\end{align}
where the reduced Green function~(\ref{redGreen}) excludes the reference
state for which the denominator would diverge to obtain
\begin{equation}
\Delta E_{\rm AC,0}=-e^2 {\cal E}_{\rm DC}^2
\Big\langle \phi\Big|z
\Big(\frac{1}{H_0-E_\phi}\Big)'
z\Big|\phi\Big\rangle=\Delta E_{\rm DC},
\end{equation}
which is the expression for the second order DC Stark shift. The static
electric field strength ${\cal E}_{\rm DC}$ is matched to the harmonic
laser field $E_L(t)={\cal E}_{\rm L}\cos(\omegaL t)$ by averaging the
laser field strength squared over one optical period:
\begin{equation}
\overline{E^2_{\rm L}(t)}=\frac12{\cal E}^2_{\rm L}\to
{\cal E}_{\rm DC}^2.
\end{equation}

\end{document}